\documentclass[12pt, amssymb]{article} %Latex2e

\usepackage[dvips]{graphicx}            % to include images
\input scrload
\newcommand{\newsection}[1]{
\addtocounter{section}{1} \setcounter{equation}{0}
\setcounter{subsection}{0} \addcontentsline{toc}{section}{\protect
\numberline{\arabic{section}}{{\rm #1}}} \vglue .6cm \pagebreak[3]
\noindent{\bf  \thesection. #1}\nopagebreak[4]\par\vskip .3cm}
\newcommand{\newsubsection}[1]{
\addtocounter{subsection}{1}
\addcontentsline{toc}{subsection}{\protect
\numberline{\arabic{section}.\arabic{subsection}}{#1}} \vglue .4cm
\pagebreak[3] \noindent{\it \thesubsection.
#1}\nopagebreak[4]\par\vskip .3cm}

\renewcommand{\theequation}{\thesection.\arabic{equation}}

\newlength{\extraspace}
\newlength{\extraspaces}
\newcounter{dummy}
\newcommand{\bc}{\begin{center}}
\newcommand{\ec}{\end{center}}
\newcommand{\be}{\begin{equation}
\addtolength{\abovedisplayskip}{\extraspaces}
\addtolength{\belowdisplayskip}{\extraspaces}
\addtolength{\abovedisplayshortskip}{\extraspace}
\addtolength{\belowdisplayshortskip}{\extraspace}}
\newcommand{\ee}{\end{equation}}

\newcommand{\ba}{\begin{eqnarray}
\addtolength{\abovedisplayskip}{\extraspaces}
\addtolength{\belowdisplayskip}{\extraspaces}
\addtolength{\abovedisplayshortskip}{\extraspace}
\addtolength{\belowdisplayshortskip}{\extraspace}}
\newcommand{\ea}{\end{eqnarray}}

\newcommand{\ban}{\begin{eqnarray*}
\addtolength{\abovedisplayskip}{\extraspaces}
\addtolength{\belowdisplayskip}{\extraspaces}
\addtolength{\abovedisplayshortskip}{\extraspace}
\addtolength{\belowdisplayshortskip}{\extraspace}}
\newcommand{\ean}{\end{eqnarray*}}
\newcommand{\baa}{
\addtocounter{equation}{1} \setcounter{dummy}{\value{equation}}
\setcounter{equation}{0}
\renewcommand{\theequation}{\thesection.\arabic{dummy}\alph{equation}}
\begin{eqnarray}
\addtolength{\abovedisplayskip}{\extraspaces}
\addtolength{\belowdisplayskip}{\extraspaces}
\addtolength{\abovedisplayshortskip}{\extraspace}
\addtolength{\belowdisplayshortskip}{\extraspace}}
\newcommand{\eaa}{
\end{eqnarray}
\setcounter{equation}{\value{dummy}}
\renewcommand{\theequation}{\thesection.\arabic{equation}}}

\newcommand{\HH}{{\cal H}_{{\rm BCFT}_0}}
\newcommand{\V}{{ V}}
\newcommand{\ra}{\rangle}
\newcommand{\la}{\langle}
\newcommand{\QB}{Q}
\newcommand{\QQ}{{\cal Q}}
\newcommand{\HD}{D_H}
\begin{document}
\begin{flushright}
September 2005\\
{\tt hep-th/0509129}
\end{flushright}
\vspace{2cm}

\thispagestyle{empty}

\begin{center}
{\Large\bf  String Field Theory \\[20mm] }

{\sc Leonardo Rastelli}\\[2.5mm]
{\it
 Joseph Henry Laboratories, \\
 Princeton University, 
Princeton, NJ 08544}, USA\\
[30mm]

{\sc Abstract}

\end{center}

\noindent

This article is a concise review of covariant string field theory prepared
for the Encyclopedia of Mathematical Physics, Elsevier (2006).
Referencing follows the publisher's guidelines.

  \vfill

\newpage

\renewcommand{\Large}{\normalsize}
%\tableofcontents

\newsection{Introduction}

String field theory (SFT) 
is the second-quantized approach to
 string theory.  In the usual, first-quantized,
formulation of string perturbation theory,
 one postulates a recipe for the string S-matrix
 in terms  of a sum over two-dimensional
worldsheets embedded in spacetime. Very schematically,
\be \label{polyakov}
\langle \langle   \,  {\V}_1 (k_1) \dots 
{\V}_n(k_n)  \, \rangle \rangle  = \sum_{topologies} g_s^{- \chi} \int [d\mu_\alpha]  \, \langle    {\V}_1 (k_1) \dots 
{\V}_n(k_n)  \rangle_{ \{  \mu_\alpha \} }\,.
\ee
Here  the {\it lhs} stands for the S-matrix
of the  physical 
string states $\{ {\V}_a (k_a) \}$.
The symbol $\langle \dots \rangle_{\{ \mu_\alpha\}}$
denotes a correlation function on the 2d  worldsheet, 
which is a punctured Riemann surface of Euler number $\chi$ and given
moduli $\{ \mu_\alpha \}$. 
 In SFT, one aims to recover this standard
prescription from the Feynman rules of a second-quantized {\it spacetime} action 
$S[\Phi]$. The string field $\Phi$, the fundamental dynamical variable,
can be  thought of as an infinite dimensional array
of  spacetime fields $\{ \phi^i(x^\mu)  \}$,  one field for each basis state
 in the Fock space of the first-quantized string.

The most straightforward way to construct $S[\Phi]$ uses 
the unitary light-cone gauge. Light-cone
 SFT is an almost immediate transcription of 
Mandelstam's light-cone diagrams in a second-quantized language.
While often useful as a book-keeping device, light-cone SFT
seems  unlikely to represent a real improvement over the first-quantized approach.
By contrast, from our experience
in ordinary quantum field theory,
we should expect Poincar\'e-covariant SFTs to give
important insights  into the issues of vacuum selection,
background independence and the non-perturbative definition
of string theory.

Covariant SFT actions are well-established
for the open (Witten, 1986), closed (Zwiebach, 1993) and open/closed (Zwiebach, 1998)
{\it bosonic} string.  These theories
are based on the BRST formalism, 
where the worldsheet  variables include the $b  c$
ghosts introduced in  gauge-fixing
 the worldsheet metric to the conformal
gauge  $g_{ab} \sim  \delta_{ab}$. (An alternative approach (Hata et al.), 
based on covariantizing  light-cone SFT,  will not
be described in this article.) 
Much less is presently known for the superstring:
classical actions have been established  for  the NS sector of the
open superstring  (Berkovits, 2001) 
 and for the heterotic string (Berkovits et al., 2004). 

During the first period of intense
activity in SFT (1985-1992), the covariant
 bosonic actions were constructed  and shown to pass the basic test of
reproducing the  S-matrix (\ref{polyakov})
to each order in the perturbative expansion.
The more recent revival of the subject (since 1999)
was triggered by the realization that SFT contains non-perturbative information
as well: D-branes emerge as solitonic solutions of the
classical equations of motion in open SFT.
 We can hope that the non-perturbative string dualities will  also
be understood in the framework of SFT, 
once covariant SFTs for the superstring  are better developed.

In this article, we review the basic formalism of covariant SFT,
using for illustration purposes the simplest model -- cubic
bosonic open SFT.  We then briefly sketch the
generalization to bosonic SFTs that include closed strings.
Finally we turn  to the subjects of classical 
solutions in open SFT and  the physics of the open string tachyon.

\newsection{Open Bosonic SFT}

The standard formulation of string theory 
starts with the choice of an on-shell spacetime background where
strings propagate.  In the bosonic string, the closed string background  is described
 by  a conformal field theory  of central charge 26 (the ``matter'' CFT).
 The total worldsheet CFT is the direct sum of this matter
 CFT and of the universal ghost CFT,  of central
 charge $-26$.  To describe open strings,  we must further specify  boundary conditions
 for the string endpoints.  
 The open string background  is encoded in a Boundary CFT, 
 a CFT defined in the upper half plane, with conformal boundary conditions
 on the real axis. (See the entry on BCFT in this Encyclopedia).  In modern
 language, the choice of BCFT corresponds to specifying a D-brane state.

 In classical open SFT, we fix  the closed
 string background (the bulk CFT) and consider 
 varying the D-brane configuration (the boundary conditions).
 To lowest order in $g_s$ we can neglect the backreaction
 of the D-brane on the closed string fields, 
 since this is a quantum
 effect from the open string viewpoint. 
 Let us prepare the ground by recalling
 the standard $\sigma$-model philosophy. To describe
 off-shell open string configurations, we should allow
 for general (not necessarily conformal)
 boundary conditions. 
 We can imagine to proceed as follows: 

i) We  choose an initial open string background,
a reference BCFT that we shall
call BCFT$_0$.  For  example, a D$p$ brane
in flat 26 dimensions (Neumann boundary conditions  on $p+1$ coordinates,
 Dirichlet on $25-p$ coordinates).

ii) We  then write  a basis of  boundary perturbations around {\it this} background. 
Taking for example BCFT$_0$ to be a D25 brane in flat space, the worldsheet action
$S_{WS}$ takes the schematic form

\be \label{ws}
S_{WS} = \frac{1}{2 \pi \alpha'} \int_{\mathbf{C}} \, 
\partial X_\mu \bar \partial X^\mu    
+ \int_{\mathbf{R}} \, 
\tilde  T(x^\mu) + \tilde A_\nu (x^\mu) \partial X^\nu  +\tilde  B_\nu (x^\mu) \partial^2 X^\nu + \dots 
\ee
Here  to the standard free bulk   action
(integrated over the whole complex plane $\mathbf{C}$)
we have added 
a perturbation localized on the real axis $\mathbf{R}$.
Notice that the basis of perturbations depends on the chosen  BCFT$_0$.

iii) We interpret the coefficients $\{ \tilde \phi^i (x^\mu) \}$ 
of the perturbations as spacetime  {\it fields}.  
(The tilde on $\tilde \phi^i(x)$ serves
as a reminder that these fields
 are not quite the same as the fields $\phi^i(x)$ that
 will appear in the open SFT action).
We are after  a 
 spacetime action $S[ \{ \tilde \phi^i \} ]$  such that 
{\it solutions} of  its classical equations of motion  correspond to {\it conformal}  boundary conditions:
\be \label{beta}
\frac{\delta S}{ \delta  \tilde \phi^i} = 0 \; ({\rm spacetime})\;
    \leftrightarrow \;  \beta_i [ \{ \tilde \phi^j \}]  = 0  \; ({\rm worldsheet}) \,.
\ee
We recognize in (\ref{ws}) the familiar
open string tachyon $\tilde T(x)$ and gauge field $\tilde A_\mu(x)$, which
are the lowest modes  in an infinite tower of fields. 
Relevant perturbations on the worldsheet (with conformal dimension $h <1$)
correspond to tachyonic fields in spacetime $(m^2 < 0)$, whereas marginal
worldsheet perturbations $(h=1)$
give massless spacetime fields.
To achieve a complete description,  we must include all the higher massive open string modes as well, which
  correspond to non-renormalizable boundary perturbations $(h >1)$.  
  In the traditional $\sigma$-model approach, this appears like a daunting task.
  The formalism of open SFT will automatically circumvent this difficulty.

\newsubsection{The Open String Field}

In covariant SFT  the reparametrization ghosts play a crucial role.
The ghost CFT consists of the Grassmann odd
fields {$b(z)$, $c(z)$, $\bar b(\bar z)$, $\bar c( \bar z)$},
of dimensions  $(2, 0)$, $(-1, 0)$, $(0,2)$, $(0,-1)$.
The boundary conditions on the real axis
are $b = \bar b $, $c = \bar c$.
The state space $\HH$ 
of the full matter + ghost Boundary CFT
 can be broken  up into subspaces of definite
ghost number,
\be
\HH = \bigoplus_{G = -\infty}^\infty \HH^{(G)} \,.
\ee
We use conventions where the SL(2, $\mathbf{R}$) 
vacuum $| 0 \rangle$ carries zero ghost number,  $G (| 0 \rangle) = 0$,
while $G(c) = +1$ and $G(b) =-1$.
As is familiar from the first-quantized treatment,
 physical open string states 
are identified with  $G=+1$
cohomology classes of the BRST operator,
\be \label{phys}
\QB | V_{phys} \rangle = 0 \, , \quad | V_{phys} \rangle \sim  | V_{phys} \rangle + \QB 
| \Lambda \rangle \, ,
\quad G(| V_{phys} \rangle)= +1\, ,
\ee
where the nilpotent BRST operator $\QB$ has the standard expression
\be
\QB = \frac{1}{2 \pi i} \oint  \left( c\, T_{matter} \,+ : b c \partial c: \right) \,.
\ee
Though not {\it a priori} obvious, it turns out that
the simplest form of the open SFT action is achieved
by taking as the fundamental off-shell variable
an  {\it arbitrary} $G= +1$ element of the first-quantized
Fock space,
\be
|\Phi \rangle  \in \HH^{(1)} \,.
\ee
By the usual state-operator correspondence of CFT,
we can also represent $|\Phi \rangle$ as a local (boundary)
 vertex operator  acting on the vacuum,
\be
|\Phi \ra = \Phi(0) | 0 \ra \, .
\ee
The open string field $|\Phi \rangle$ is really an infinite-dimensional array of spacetime fields.
We can make this transparent by expanding  it  as
\be\label{expansion}
|\Phi \rangle = \sum_i  \int \,  d^{p+1}k  \;    | \Phi_i(k)  \rangle \, \phi^i(k^\mu)  \, ,
\ee
where $\{ | \Phi_i (k)\ra \}$ is some convenient basis of $\HH^{(1)}$
that diagonalizes the momentum $k_\mu$.  
The fields $ \phi^i$   are a priori complex. This is remedied by imposing
 a suitable reality condition on the string field,
 which will be stated  momentarily.
Notice there are many more elements
in $\{  |\Phi_i (k)\ra \}$  than in the physical
 subspace (the cohomology classes of $\QB$).
 Some of the extra  fields will turn out to be non-dynamical
 and could be integrated out, but at
 the price of making the OSFT action look much more complicated.
 
It is often useful to think of the string field in terms of its
Schr\"{o}edinger representation, that is, as a functional
on the configuration space of  open strings.   
Consider the unit half--disk in the upper--half plane, 
$\HD \equiv \{ |z| \leq 1, \Im  \,z \geq
0 \}$, with the vertex operator  ${\Phi}(0) $
inserted at the origin. Impose BCFT$_0$ open string boundary
conditions for the  fields $X(z, \bar z)$  on the real axis
(here $X(z, \bar z)$ is a shorthand for all matter {\it and} ghost fields), 
and  boundary conditions $X (\sigma)=X_b(\sigma)$ on the
curved boundary of $\HD$, $z =\exp(i \sigma)$, $0 \leq \sigma \leq \pi$.
 The path-integral over $X(z, \bar z)$ in the interior of
 the half--disk assigns a complex number
 to any given  $X_b(\sigma)$,
 so we obtain a functional $\Phi[X_b(\sigma)]$.
This is the Schr\"{o}edinger wavefunction 
of the state $ \Phi(0) | 0 \ra $.   Thus we can think of
 open string functionals $\Phi[X_b(\sigma)]$
as the fundamental variables of OSFT. This is
as it should be: the first-quantized wavefunctions are promoted
 to dynamical fields  in the second-quantized theory.
Finally let us quote the reality condition for the string field,
which takes a compact form in the Schr\"{o}edinger representation:
\be
\Phi[ X^\mu (\sigma), b(\sigma), c(\sigma) ] = \Phi^*[ X^\mu (\pi -\sigma), b(\pi -\sigma), c(\pi -\sigma) ] \, ,
\ee
where the superscript $*$ denotes complex conjugation. 

\newsubsection{The Classical Action}

With all the ingredients  in place, it is immediate to
write the {\it quadratic} part of the OSFT action. The linearized
equations of motion must reproduce the physical-state
condition (\ref{phys}).  This suggests
\be \label{S2}
S  \sim  \la \Phi | \, \QB \, | \Phi \rangle \,.
\ee
Here $\la  \; |\;  \ra$ is the usual BPZ inner product
of BCFT$_0$, which 
is defined in terms of a two-point correlator on the disk,
as we review below. The ghost anomaly implies 
 that on disk we must have $G_{tot} = +3$, which   happily  is the case in  (\ref{S2})
Moreover since the inner product is non-degenerate, variation of (\ref{S2})
gives
\be
\QB |\Phi \ra = 0 \,  ,
\ee
as desired. The equivalence relation $|V_{phys} \rangle \sim |V_{phys} \rangle+ \QB |\Lambda\rangle$
is interpreted in the second-quantized language
as  the {\it spacetime} gauge-invariance
\be \label{gauge}
\delta_\Lambda |\Phi \rangle = \QB | \Lambda \rangle \, ,  \quad   | \Lambda \ra \in \HH^{(0)} \, ,
\ee
valid for the general off-shell field.  This equation
is a very compact generalization of the linearized gauge-invariance
for the massless gauge-field. Indeed,
 focusing on the  level-zero components, $ | \Phi \rangle  \sim  A_\mu (x)\,  (c \partial  X^\mu )(0) | 0 \rangle$
and $|\Lambda \rangle \sim \lambda(x) |0 \rangle$,
we find $\delta A_\mu (x)= \partial_\mu \lambda(x)$.
It is then plausible to guess that the non-linear gauge-invariance
should take the form
\be \label{nonlinear}
 \delta_\Lambda |\Phi \rangle = \QB | \Lambda \rangle  + |\Phi \ra * | \Lambda \ra - |\Lambda \ra * | \Phi \ra\, ,
\ee
where $*$ is some suitable product operation that conserves ghost number
\be
*  \, : \; \HH^{(n)}   \otimes \HH^{(m)}   \to \HH^{(n+m)} \,.   
\ee
Based on a formal analogy with 3d non-abelian Chern-Simons theory,
Witten proposed the cubic action
 \be \label{OSFT}
S = -\frac{1}{g_o^2} \left( \frac{1}{2} \la \Phi | \QB | \Phi \ra +
\frac{1}{3} \,\la \Phi |  \Phi * \Phi \ra \right)\,  .
\ee
The string field $| \Phi \ra$ is analogous to
 the Chern-Simons gauge-potential $A =A_i dx^i$, 
 the $*$ product to the $\wedge$ product
 of differential forms, $\QB$ to the exterior derivative $d$,
 and the ghost number $G$ to the degree of the form.  
The analogy also suggests a number
of algebraic identities: 
\ba \label{algebraic}
Q^2 & = & 0 \, , \\
\la Q \, A \, | \, B \ra  & = &  - (-1)^{G(A)} \la A \, | \, Q\, B \ra \, , \nonumber \\ 
Q(A * B) & = &  (Q\, A) * B + (-1)^{G(A)} A * (Q\, B) \, , \nonumber \\
  \la   A \,   | \,  B  \ra &  =  & (-1)^{G(A) G(B)}   \la B \, | \, A \ra \, , \nonumber \\
    \la A \, | \, B * C \ra   & =  &   \la  B\, | \,  C * A \ra  \, ,\nonumber \\
A * (B * C)  & =&  (A * B) * C \, . \nonumber
\ea
Note in particular the {\it associativity} of the $*$-product.
It is straightforward to check that this algebraic
structure implies the gauge-invariance of the cubic action under
(\ref{nonlinear}).  
A $*$-product satisfying all
required formal properties can indeed be defined.
The most intuitive presentation is in the functional language.

\begin{figure}[th]
\begin{center}
       % \resizebox{\textwidth}{!}{
            \scalebox{0.7}{
               \includegraphics[width=\textwidth]{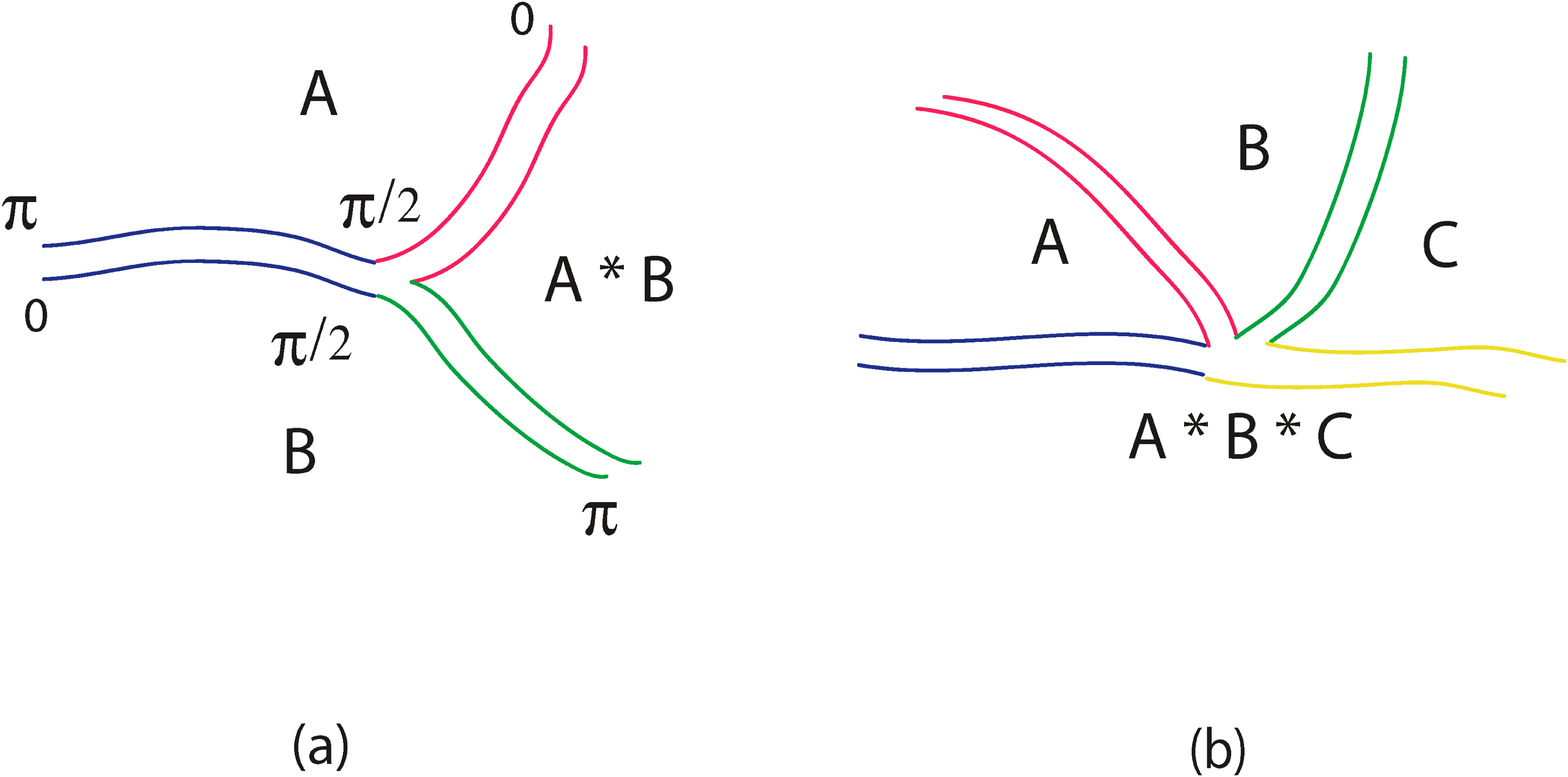}
               }
                \end{center}
                \vspace{-1cm}
                \caption{ \it Midpoint overlaps of open strings.} \label{star}
 \end{figure}

Given an open string curve $X(\sigma)$, $0 \leq \sigma \leq  \pi$,
we single out the string midpoint $\sigma = \pi/2$ 
and define the left and right ``half-string'' curves 
\ba
X_L(\sigma) &  \equiv & X(\sigma) \,   \qquad \quad {\rm for} \;  \; 0 \leq \sigma \leq \frac{\pi}{2} \, , \\
X_R(\sigma)&  \equiv & X(\pi -\sigma) \, \quad {\rm for} \; \; \frac{\pi}{2} \leq \sigma \leq \pi \,. \nonumber
\ea
A functional $\Phi[X(\sigma)]$ can of course be regarded
as a functional of the two half-strings, $\Phi[X] \to \Phi[X_L, X_R] $.
We define 
\be \label{functional}
 (\Phi_1 * \Phi_2)[X_L, X_R]  \equiv \int [dY] \,    \Phi_1[X_L, Y]  \Phi_2[Y, X_R] \, ,
 \ee
 where $\int [dY]$ is meant as the functional integral
 over the space of half-strings  $Y(\sigma)$,  with $Y(\pi/2) = X_L(\pi/2) = X_R (\pi/2)$.
 The picture (Fig.\ref{star}a) is that of two open strings interacting 
(to form a single open string) if and only if the right half of the first string
 precisely overlaps with the left half of the
 second string. Associativity is transparent (Fig.\ref{star}b).

%%%%

\begin{figure}[th]
\begin{center}
        %\resizebox{\textwidth}{!}{
            \scalebox{0.8}{
               \includegraphics[width=\textwidth]{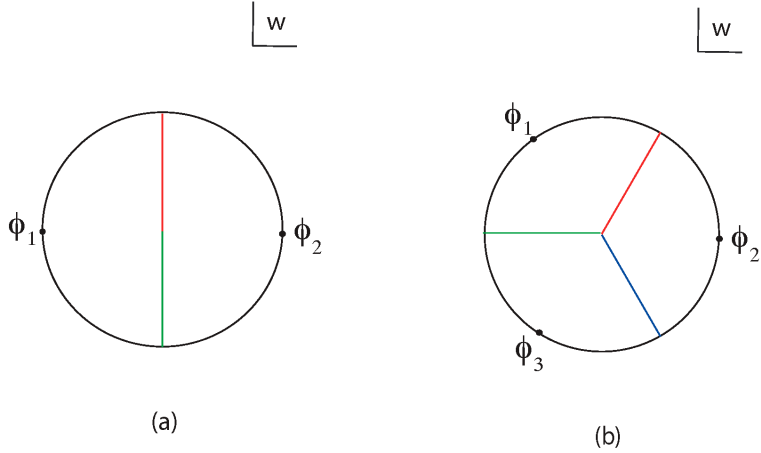}
              % \vspace{-3cm}
                }
              \vspace{0cm}  
              \end{center}
                \caption{ \it Representation of the quadratic and
cubic vertices as 2-- and 3--punctured
unit disks.} \label{D}
 \end{figure}

  We can now translate this formal construction
  in the precise CFT language.   Very generally,  an $n$-point vertex of open strings
  can be defined by specifying an $n$-punctured disk, 
  that is, a disk with marked points on the boundary (punctures)
  and a choice of local coordinates 
around each puncture. Local coordinates
are essential since we are dealing with off-shell  open string states.
The BPZ inner product (2pt vertex) is given by 
  \be
 \la \Phi_1 | \Phi_2 \ra \equiv \la I \circ \Phi_1(0)\,\, \Phi_2(0) \ra_{\rm UPH}\,,  \quad  I(z) = -\frac{1}{z} \,.
\ee
The symbol $f \circ \Phi(0)$, where $f$ is a complex map, means the conformal transform
of  $\Phi(0)$ by $f$. For example if  $\Phi$ is a dimension $d$ primary
field, then  $f \circ \Phi(0) = f'(0)^d \Phi(f(0))$. If $\Phi$ is non--primary
the transformation rule will be more complicated and involve
extra terms with higher derivatives of $f$.  By performing the SL(2, $\mathbf{C}$) transformation 
\be
w = h(z) \equiv \frac{1+iz}{1-iz} 
\ee
we can represent the 2pt vertex as a correlator
on the unit disk {$D = \{ |w| \leq 1 \}$},
\be
 \la \Phi_1 | \Phi_2 \ra  = \langle   f_1 \circ \Phi_1(0) \,  , \;  f_2 \circ \Phi_2(0) \rangle_{ D} \, ,  \quad  f_1(z_1) = -h(z_1)\, , \; f_2(z_2) = h(z_2) \, .
 \ee
 The vertex operators are inserted as $w=-1$ and $w=+1$ on $D$ (see Fig.\ref{D}a)
 and correspond to the two open strings at  (Euclidean) worldsheet time $\tau = -\infty$
 (we take $z =\exp(i \sigma + \tau)$).  The left half of $D$ is the worldsheet of the
 first open string; the right half of $D$ is the worldsheet
 of the second string. The two strings meet at $\tau = 0$ on the imaginary $w$ axis.
The 3pt Witten vertex is given by 
\be \label{cubic}
 \la \Phi_1 , \Phi_2 , \Phi_3\ra \equiv 
\la g_1 \circ \Phi_1(0) g_2 \circ \Phi_2(0) 
 g_3 \circ \Phi_3(0) \ra_{D}\, ,
 \ee
where
\be
 \label{fi}
g_1(z_1)  =   e^{\frac{2 \pi i}{3}}\left(
\frac{1+iz_1}{1-iz_1} \right)^{\frac{2}{3}}\,, \;
  g_2(z_2)   = \left(
\frac{1+iz_2}{1-iz_2} \right)^{\frac{2}{3}}\,, \;
  g_3(z_3)   = e^{-\frac{2 \pi i}{3}} \left(
\frac{1+iz_3}{1-iz_3} \right)^{\frac{2}{3}}\, . \nonumber
\ee
The three punctured disk is depicted in Fig.\ref{D}b,
and describes the symmetric midpoint overlap of the three
strings at $\tau = 0$.  Finally the relation between the 3pt vertex and the $*$-product is 
\be \label{trick}
\la \Phi_1  |  \Phi_2 * \Phi_3 \ra \equiv \la \Phi_1, \Phi_2, \Phi_3 \ra \,.
\,    \ee
Knowledge of the {\it rhs} in (\ref{trick}) for all $\Phi$ allows to reconstruct the $*$-product.
All formal properties (\ref{algebraic})  are easily shown to
hold in the CFT language. This completes the definition of the OSFT action.
        
     Evaluation of the classical action is completely algorithmic and can
be carried out for arbitrary massive
states, with no fear of divergences,
since in all required correlators
the operators are inserted well apart from each other.

\newsubsection{Quantization}

Quantization is defined by the  path-integral over the second-quantized
  string field.  The first step is to deal with the gauge-invariance
(\ref{nonlinear}) of the classical action. 
The gauge symmetry is reducible:
not all gauge-parameters $\Lambda^{(0)}$
(the superscript labels ghost number) lead to a
gauge transformation. This is clear
at the linearized level, indeed, if $\Lambda^{(0)} = \QB \Lambda^{(-1)}$,
then $\delta_{\Lambda^{(0)} }\Phi^{(1)} =\QB^2 \Lambda^{(0)}= 0$. 
Thus the set  $\{ \Lambda^{(0)} \}$ gives a redundant parametrization
of the gauge group. Characterizing
this redundancy is somewhat  subtle, since 
fields of the form $\Lambda^{(-1)} = \QB \Lambda^{(-2)}$
do not really lead to a redundancy in $\Lambda^{(0)}$, 
and so on, {\it ad infinitum}.
It is clear that we need  to introduce an infinite
tower of (second-quantized) ghosts for ghosts.

The Batalin-Vilkovisky  formalism  is a  powerful
way to handle the problem.   The basic object  
is the master action $S(\phi^s, \phi_s^*)$,
which is a function of the ``fields'' $ \phi^s$ and of the
 ``antifields'' $\phi_s^*$.  Each field is paired with
 a corresponding antifield of opposite Grassmanality.
The master action must obey the boundary condition of  reducing to
 the classical action when the antifields are set to zero.
(Note  that in general the set of fields
$ \phi^s $ will be larger than the set of fields $\phi^i$ that appear
in the classical action). Independence of the S-matrix on the gauge-fixing
procedure is equivalent to the
 BV {\it master equation} 
\be \label{BV}
\frac{1}{2} \{ S, S\} = -\hbar \, \Delta S \, .
\ee
The antibracket $\{ \; ,  \; \}$ and the $\Delta$ operator
are defined as
\be
\{ A, B \} \equiv   \frac{\partial_r A}{\partial \phi^s}  \frac{\partial_l B}{\partial \phi_s^*} -   \frac{\partial_r A}{\partial \phi_s^*}  \frac{\partial_l B}{\partial \phi^s}\, ,\qquad
\Delta \equiv \frac{\partial_r}{\partial \phi^s} \frac{\partial_l}{\partial \phi_s^*} \, ,
\ee
where $\partial_l$ and $\partial_r$ are derivatives
from the left and from the right.  It is often
convenient to expand $S$ in powers of $\hbar$,
$S = S_0 + \hbar S_1 + \hbar^2 S_2 + \dots$, with 
\be
\{ S_0, S_0\} = 0 \, ,  \quad \{S_0, S_1 \} + \{S_0, S_1 \}  = -2 \hbar \Delta S_0 \,, \quad  \dots
\ee
With these definitions in place, we shall simply
describe the answer, which is   extremely elegant.
In OSFT the full set of fields and antifields is packaged
in a single string field $|\Phi\rangle$, of {\it unrestricted}
ghost number. If we write
\be
|\Phi \rangle= |\Phi_- \rangle + |\Phi_+  \rangle \, ,\quad  {\rm with} \; G(\Phi_- ) \leq 1 \; {\rm and} \; G(\Phi_+) \geq 2 \,   ,
\ee
all the  fields are contained in $|\Phi_- \rangle $ and all the
antifields in $|\Phi_+ \rangle$. To make the pairing
explicit, we pick a basis $\{ |\Phi_s \ra \}$
of $\HH$, and define a conjugate basis  $\{  |   \Phi_s^C \ra \} $ by
\be
\langle \Phi^C_r | \Phi_s \rangle = \delta_{rs} \, .
\ee
Clearly $G( \Phi^C_s ) + G( \Phi_s ) =3 $. 
Then 
\be
| \Phi_- \rangle = \sum_{G (  \Phi_s ) \leq 1} | \Phi_s \rangle \, \phi^s \; ,\qquad
| \Phi_+ \rangle = \sum_{G (  \Phi_s ) \leq 1} | \Phi^C_s \rangle \, \phi_s^* \,.
\ee
Basis states $| \Phi_s  \ra$ with even (odd) ghost number
$G(\Phi_s)$ are defined to be Grassmann even (odd).
The full string field  $|\Phi\rangle$ is declared to be Grassmann odd.
It follows that   $\phi^s$ is Grassmann even (odd) for $G(\Phi_s)$ odd (even),
and that the corresponding antifield  $\phi_s^*$
has the opposite Grassmanality of $\phi^s$, as it must be.
With this  understanding of $| \Phi \rangle$, the  classical master action
$S_0$ is identical in form to the Witten action (\ref{OSFT})!
The boundary condition is satisfied, indeed  setting $| \Phi_+ \rangle = 0$
the ghost number anomaly implies that  only the  terms
with $G = +1$ survive.
The equation  $\{ S_0 , S_0\} = 0$ follows from straightforward manipulations using
the algebraic identities (\ref{algebraic}). On the other hand,
the issue of whether $\Delta S_0 = 0$,
or whether instead quantum corrections are needed to  satisfy full BV master equation,   
is more subtle and has never been fully resolved.
 The  $\Delta$ operator receives singular 
 contributions from the same region
 of moduli space responsible for the appearance
 of closed string poles, discussed below in section 2.5.
 (See (Thorn, 1989) for the classic statement of this issue).
 It seems possible to choose of basis in $\HH$ 
 such that there are no quantum corrections to $S_0$
 (Erler and Gross, 2004). In the following we shall derive the Feynman rules
implied by $S_0$ alone.

\newsubsection{SFT Diagrams and Minimal Area Metrics}

Imposing the Siegel gauge condition $b_0 \Phi = 0$,
one finds the gauge-fixed action
\be
 S_{gf}  = -\frac{1}{g_o^2} \left( \frac{1}{2} \la \Phi | \, c_0 L_0 \Phi \ra +
\frac{1}{3} \,\la \Phi |  \Phi * \Phi \ra    + \la \beta |\, b_0 | \Phi \ra \,  \right)\,  ,
\ee
where $\beta $ is a Lagrangian multiplier.
The propagator reads
\be
\frac{b_0}{L_0} = b_0 \, \int_0^\infty  dT \, e^{-T L_0} \,.
\ee
Since $L_0$ is the first-quantized open string Hamiltonian,
$e^{-T L_0}$ is the operator that evolves
the open string wave-functions $\Psi [ X(\sigma)]$
by Euclidean worldsheet time $T$. It can be
visualized as a flat rectangular strip of ``horizontal'' width $\pi$ and ``vertical''  height $T$.
Each propagator comes with an antighost insertion
\be \label{insertion}
b_0 = \int_0^\pi b(\sigma)
\ee
integrated on a horizontal trajectory.   
%%%%%
\begin{figure}[th]
\begin{center}
        %\resizebox{\textwidth}{!}{
            \scalebox{0.9}{
               \includegraphics[width=\textwidth]{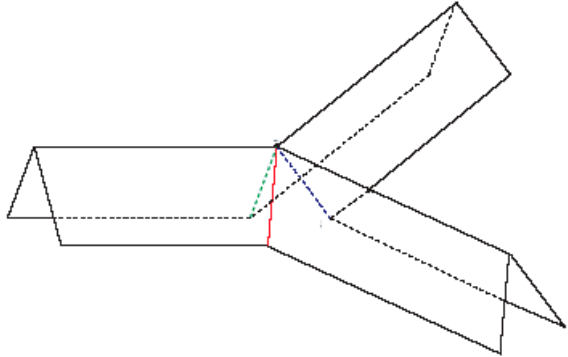}
              % \vspace{-3cm}
                }
                \end{center}
                \vspace{-1.5cm}
                \caption{ \it The cubic vertex represented as the midpoint
gluing of three strips.} \label{strips}
 \end{figure}
%%%%

The only elementary interaction vertex is the midpoint
three string overlap, visualized in Fig.\ref{strips}.
We are instructed to draw all possible diagrams
with given external legs (represented as semi-infinite strips),
and to integrate over
all  Schwinger parameters $T_i \in [0, \infty)$ 
associated with the internal propagators.
The claim is that  
this prescription reproduce precisely the first-quantized  
result  (\ref{polyakov}).   This follows if we can show that (i) the OSFT Feynman  rules
   give a unique cover of the  moduli space
of open Riemann surfaces; (ii) the integration measure
 agrees with the measure $[d \mu_\alpha]$ 
in (\ref{polyakov}). 
This latter property  holds  because  the antighost insertion (\ref{insertion})
 is precisely the one prescribed
 by the Polyakov formalism  for integrating
 over the moduli $T_i$.
To show point (i), we  introduce the 
 concept of minimal area metrics, which  has  proven very  fruitful. 
 (See  Giddings et al. (1986), Giddings (1988) for the open SFT
 case and  Zwiebach (1993, 1998) for the closed and open/closed SFTs).
 Quite generally,
 the Feynman rules of a SFT provide us with a cell decomposition
 of the appropriate moduli space of Riemann surfaces,
 a way to construct surfaces in terms
 of vertices and propagators.  Given a Riemann surface
 (for fixed values of its complex moduli), the SFT must
 associate with it one and only one string diagram.
 The diagram has more structure than the Riemann
 surface: it defines a metric on it.
 In all known covariant SFTs, this is the metric of {\it minimal
 area}  obeying suitable length conditions.
Consider the following

    {\bf Minimal Area problem for open SFT}.
  Let ${\cal R}_{o}$ be a Riemann surface
  with at least one boundary component
  and possibly punctures  on the boundary.
  Find the (conformal)  metric of minimal area on ${\cal R}_{o}$
  such that all non-trivial Jordan open curves have length
  greater or equal to $\pi$. (A curve
  is said to be non-trivial if it cannot be continuously shrunk
 to a point without crossing a puncture.)
 
 An OSFT  diagram (for fixed values of its $T_i$),
 defines a  Riemann surface ${\cal R}_{o}$ 
 endowed with a metric solving
 this minimal area problem. 
 This is the metric  implicit in its picture: flat everywhere except at the
 conical singularities of defect angle $(n -2) \pi$
 when $n$ propagators meet symmetrically.
 (For $n=3$, these are the elementary cubic vertices;
 for $n > 3$, they  are effective vertices, obtained
 when propagators joining  cubic vertices collapse to zero length.)
 It is not hard to see both that the length conditions
 are obeyed, and that the metric cannot be made smaller without violating
 a length condition. Conversely, any
  surface ${\cal R}_{o}$  endowed with a minimal
 area metric,  corresponds to an OSFT diagram.
 The idea is that the minimal area metric
 must have open geodesics  (``horizontal trajectories'')
 of length $\pi$  foliating the surface. The geodesics intersect on 
 a set of measure zero -- the ``critical graph'' where the propagators
 are glued. Bands of open geodesics of infinite height are
 the external legs of the diagram,  while bands of finite height are the internal propagators.

The single cover of moduli space is then
ensured by an existence and uniqueness theorem for  metrics solving the
minimal area problem for OSFT.   
These metrics are seen to arise from Jenkins-Strebel quadratic differentials.
Existence shows 
  that the Feynman rules of OSFT generate
   each  Riemann surface ${\cal R}_{o}$  at least once.
   Uniqueness shows there is no overcounting: since different
 diagrams correspond to different metrics (by inspection
 of their picture),  no Riemann surface can be generated twice.

 \newsubsection{Closed Strings in OSFT}
 
 As is familiar, the open string S-matrix 
 contains poles due to the exchange of on-shell open {\it and} closed
 strings. The closed string poles
 are present in non-planar loop amplitudes.
 We have seen that OSFT reproduces
 the standard S-matrix.
 Factorization over the open string poles
is manifest,  it corresponds
to propagator lengths $T_i$ going to infinity.
Surprisingly, the closed string poles
are also correctly reproduced, 
despite the  fact that OSFT treats only the open strings 
 as fundamental dynamical variables.
In some sense, closed strings must be considered
as derived objects in OSFT. 
 Factorizing
the amplitudes over the closed string poles
one finds that
 on-shell closed string states 
 can be represented, at least formally, as  certain singular open string fields with $G=+2$,
 closely related to the (formal)  identity string field. 
The picture is that of a folded open string,
 whose left and right halves precisely overlap,  
 with an extra closed  string vertex operator inserted at the midpoint.
The corresponding open/closed vertex 
 is given by
 \be \label{oc}
  \langle  \Psi_{phys}  | \Phi \rangle_{OC} \equiv  \langle   \Psi_{phys }(0) \,  {\cal I} \circ \Phi(0) \rangle_D \, , \quad {\cal I} =\left( 
  \frac{1+i z}{1-iz} \right)^2 \, ,
 \ee
and describes the coupling to the open
 string field of  a non-dynamical,  {\it on-shell} closed string $| \Psi_{phys} \rangle$.
 It is possible to add this open/closed vertex to the OSFT action.
Remarkably, the resulting Feynman rules give a single cover of the moduli space of Riemann surfaces
with at least one boundary, with open {\it and} closed punctures.
This is shown using the same minimal area problem as above,
but now allowing for surfaces with  closed punctures as well.

We should finally mention that
the structure of OSFT emerges frequently in topological string theory,
in contexts where open/closed duality plays a central role.
Two examples are the interpretation of Chern-Simons theory as the OSFT
for the A-model on the conifold, and the intepretation of the Kontsevich 
matrix integral for topological gravity as the OSFT on FZZT branes
in  $(2,1)$ minimal string theory.

\newsection{Closed Bosonic SFT}

The generalization to covariant closed SFT is non-trivial,
essentially because the requisite closed string decomposition
of  moduli space is much more complicated. 

The free theory parallels the open
case, with  a minor complication in
the treatment of the CFT zero  modes. 
The closed string field is taken to
live in a  {\it subspace} of the matter + ghost state  space,
 $| \Psi \ra \in \tilde {\cal H}_{{\rm CFT}_0}$,
where the tilde means that  we impose the subsidiary conditions
 \be \label{subsi}
b_0^- | \Psi \ra =L_0^- | \Psi \ra = 0\,, \quad b_0^{-} \equiv b_0 - \bar b_0 \, , L_0^{-} \equiv L_0 - \bar L_0 \, .
 \ee
  In the classical theory
the string field carries ghost number $G = +2$,  since it is the off-shell
extension of the familiar closed string physical states,
and
 the quadratic  action reads
\be \label{quadrclosed}
S  \sim  \la \Psi, Q_c  \Psi \rangle \, .
\ee
Here $Q_c $
 is the usual closed
BRST operator.  The inner product $\la \, , \ra$ 
is defined in terms of the BPZ inner product, with an extra  insertion of $c_0^- \equiv c_0 - \bar c_0$,
\be
\la A, B \rangle \equiv \la A | \, c_0^- \, |B \rangle \,. 
\ee
In (\ref{quadrclosed}) $G_{top}=+6$,
as it should be. Without the extra ghost insertion and the subsidiary
conditions (\ref{subsi}) it would not be possible to write a quadratic
action. The linearized equations of motion and gauge invariance,
 \be
 Q_c | \Psi \rangle = 0 \, , \quad |\Psi \rangle \sim |\Psi \rangle + Q_c |\Lambda \rangle \, , \quad | \Lambda \rangle  \in \tilde {\cal H}_{{\rm CFT}_0}^{(1)}\, ,
 \ee
give the expected cohomological problem. The fact
that the cohomology is computed in the
{\it semi-relative} complex,
 $ b_0^- | \Psi \rangle = b_0^- | \Lambda \rangle = 0$, 
well-known from the operator formalism of the first-quantized
theory, is recovered  naturally in the second-quantized treatment.

The interacting action is constructed
iteratively, by demanding that the
resulting Feynman rules give a (unique) cover
of moduli space. This requires the introduction of
infinitely many elementary  string vertices ${\cal V}_{g,n} $,
where $n $ is the number of closed string punctures and $g$ the genus. 
This decomposition of moduli space
is more intricate than the decomposition that arises in OSFT,
but is in fact  analogous to it, when characterized in terms of  the following

{\bf  Minimal Area problem for closed SFT}.
  Let ${\cal R}_{c}$ be a closed Riemann surface,
  possibly with punctures.
  Find the (conformal) metric of minimal area on ${\cal R}_{}$
  such that all non-trivial Jordan closed curves have length
  greater or equal to $2\pi$. 

The minimal area metric induces a foliation
of   ${\cal R}_{c}$ 
 by closed geodesics of length $2 \pi$.
In the classical theory $(g =0)$, the minimal
area metrics arise from Jenkins-Strebel quadratic
differentials (as in the open case), and geodesics intersect on a measure zero set.  
For $g > 0$ however there can be foliation bands of geodesics that cross. 
By staring at the foliation we can break up  
the surface into vertices and propagators.  In correspondence with each puncture  there is a band of infinite height,
 a flat semi-infinite cylinder of circumference $2 \pi$, which we identify as
  an external leg of the diagram. We mark a closed geodesic
  on each semi-infinite cylinder, at a distance $\pi$ from its boundary.
  Bands   of finite height   (internal bands not associated to punctures) correspond
  to propagators if their height is greater than $2 \pi$,
  otherwise they are considered part of an elementary vertex.
Along any internal cylinder of height greater than $ 2 \pi$,
 we mark two closed geodesics, at
a distance $\pi$ from the boundary of the cylinder. 
If we now cut open  all the marked curves, the surface
decomposes into a number of semi-infinite cylinders (external
legs), finite cylinders (internal propagators)
and surfaces with boundaries (elementary interactions).
Each elementary interaction of genus $g$ and with $n$ boundaries
is an element of ${\cal V}_{g,n}$.
A crucial  point of this construction
is that we took care of leaving
 a ``stub'' of length $\pi$ attached
 to each boundary. Stubs ensure that sewing
 of surfaces preserves the length condition on the metric (no closed curve
 shorter than $2 \pi$).

These geometric data can be translated
 into an iterative algebraic construction of the 
full quantum action $S[\Psi]$.
The ${\cal V}_{g, n}$ satisfy   geometric recursion
relations whose algebraic counterpart is the
quantum BV master equation for  $S[\Psi]$.
Remarkably, the singularities of the $\Delta$ operator
encountered in OSFT are absent here,
precisely because of the presence of the stubs.
We refer to (Zwiebach, 1993) for a complete discussion of closed SFT.

\newsubsection{Open/closed SFT}

There is also a covariant SFT that includes
both open and closed strings as fundamental
variables. The Feynman rules arise
from the natural

{\bf Minimal Area problem for open/closed SFT}.
  Let ${\cal R}_{oc}$ be a Riemann surface,
  with or without boundaries, 
  possibly with open and closed punctures.
  Find the (conformal) metric of minimal area on ${\cal R}_{oc}$
  such that all non-trivial Jordan  open curves have length
  greater or equal to $l_o = \pi$, and all non-trivial
Jordan  closed curves have length greater or equal to $l_c = 2 \pi$.

The surface  ${\cal R}_{oc}$ is decomposed in
terms of elementary vertices $ {\cal V}_{b,m}^{g,n}$ 
(of  genus $g$,   $b$ boundary components,
$n$ closed string punctures  and $m$ open string punctures)
joined by open and closed propagators.
Degenerations of the surface correspond
always to propagators becoming of infinite
length -- factorization is manifest both
in the open and in the closed channel.

The  SFT  described in section 2.5 (Witten OSFT
augmented with the single open/closed vertex (\ref{oc}))
corresponds to taking $l_o = \pi$ and
$l_c = 0$. 
 Varying $l_c  \in [0,  2 \pi]$,
we find a whole family of interpolating SFTs. This
construction clarifies the special status
of the Witten theory:  moduli space is 
covered by a single cubic open overlap vertex, with no need to introduce
 dynamical closed strings, but at the price of a somewhat singular
formulation.

\newsection{Classical Solutions in Open SFT}

In the present formulation of  SFT, 
a  background  (a classical
solution of string theory) 
must be chosen from the outset. The
very definition of the string field requires to specify a (B)CFT$_0$.
 Intuitively,   the string field  lives in the ``tangent'' 
to theory space at a specific point   -- where ``theory space'' is some notion of
a ``space of 2d (boundary) quantum field theories'',
not necessarily conformal. 
In the early 1990's   independence
on the choice of background
 was demonstrated  for {\it infinitesimal}
deformations:   the SFT actions written using neighboring (B)CFTs
    are indeed related by a field  redefinition.
In recent years it has become apparent that 
at least
the {\it open} string field reaches out to open string  backgrounds a {\it finite} distance
away --  possibly covering
 the whole of  theory space.
(Classical solutions of closed SFT are beginning to be investigated
at the time of this writing (2005)).  

The OSFT action written using  BCFT$_0$ data
is 
 just the full worldvolume action  of the D-brane 
with BCFT$_0$ boundary conditions. 
Which classical solutions should we expect in
this OSFT? In the bosonic string,
 D$p$ branes carry no conserved charge and are unstable.
This  instability is reflected in the presence
 of a mode with $m^2  = -1/\alpha'$, 
the open string tachyon $T(x^\mu)$, $\mu=0, \dots p$. From this physical picture, 
Sen argued that:

 (i) The tachyon potential, 
 obtained by eliminating
the higher modes of the string field by their
equations of motion,   must admit a local minimum
 corresponding to the vacuum with no D-brane at all (henceforth,
the {\it tachyon vacuum}, $T(x^\mu) = T_0$).  

(ii) The value of the potential at $T_0$ (measured with respect
to the BCFT$_0$ point $T=0$) must be exactly equal
to minus the tension of the brane with BCFT$_0$ boundary conditions.

(iii) There must be no perturbative open string
excitations around the tachyon vacuum. 

(iv)There must be space-dependent ``lump'' solutions 
corresponding to lower-dimensional branes. For example
a lump localized along one worldvolume direction, say $x^1$,
such that $ T(x^1) \to T_0$  as  $x^1 \to \pm \infty$,
is identified with a D$(p-1)$ brane.

Sen's conjectures
have all been verified in OSFT. (See (Sen, 2004) and  (Taylor and Zwiebach, 2003) for reviews).
The  deceptively simple-looking equations of motion (in Siegel gauge) 
\be
L_0 | \Phi \ra +b_0( | \Phi \ra * | \Phi \ra) = 0 \, ,
\ee
are really an infinite system of coupled equations, and no analytic solutions are known.
Turning on a vev for the tachyon drives into condensation
an infinite tower of modes.
Fortunately the approximation technique of ``level truncation''
is surprisingly effective. The string field
is restricted to modes with an $L_0$ eigenvalue smaller than a prescribed
maximal level $L$. For any finite $L$, the truncated OSFT  contains
a finite number of fields and numerical
computations are possible. Numerical results for various classical
solutions converge quite rapidly as the level $L$ is increased.

The most important solution is the string field
 $|{\cal T}\ra$ that corresponds to the tachyon vacuum.
A remarkable feature of $|{\cal T}\ra$ is {\it universality}:
it 
can be written as a linear combination
of modes obtained by acting on the tachyon $c_1|0 \ra$ with
 ghost oscillators and matter Virasoro operators,
$$
|{\cal T} \ra= T_0  \, c_1 | 0 \ra  + u \, L^m_{-2} c_1 | 0 \ra +
v \, c_{-1} | 0 \ra + \dots
$$
This implies that the properties of $|{\cal T}\ra$ are independent
of any detail of BCFT$_0$, since
all computations involving $|{\cal T}\ra$ 
 can be reduced to purely combinatorial manipulations 
 involving the ghosts and the Virasoro algebra.
The numerical results strongly confirm Sen's conjectures,
and indicate that the tachyon vacuum is 
located at a non-singular point in configuration space. 
Numerical solutions describing lower--dimensional
branes and exactly marginal deformations are also
available.  For example, the full family
of solutions interpolating between a D1 and a D0 brane
at the self-dual radius has been found. There is increasing
evidence that the open string field provides
a faithful map of the open string landscape.

\newsubsection{Vacuum SFT: D-branes as Projectors}

In the absence of  a closed-form expression for $| {\cal T} \ra$,
we are led to guesswork. When expanded
around  $| {\cal T} \ra$, the OSFT is still cubic,
only with a different kinetic term ${\cal Q}$,
\be \label{vsft}
S = -\kappa_0 \Big[ {1\over 2} \langle \Phi \, | \QQ \, | \Phi\rangle +
{1\over 3} \langle \Phi \, | \, \Phi * \Phi \rangle \Big]\, .
\ee
The operator ${\cal Q}$ must obey
all the formal properties (\ref{algebraic}), must
 be universal (constructed from ghosts
and matter Virasoro operators),  and must have
trivial cohomology  at $G= +1$.
Another constraint comes from
requiring that (\ref{vsft}) admits classical
solutions in Siegel gauge. The choice 
\be \label{QQ}
\QQ ={1\over 2i} (c(i) - \bar c(i)) = 
c_0 - (c_2 + c_{-2}) +  (c_4 + c_{-4}) - \cdots \, ,
\ee
satisfies all these requirements.  The conjecture (Rastelli, Sen and Zwiebach, 2001)
is that by a  field redefinition, the  kinetic term around the tachyon vacuum can be cast into this
form. This ``purely ghost'' $\QQ$ is somewhat 
singular (it acts  at the delicate string midpoint),
 and presumably should be regarded as the leading term  
 of a more complicated operator that includes
matter pieces as well. The normalization constant
$\kappa_0$ is formally infinite. Nevertheless,
 a regulator (for example level truncation) 
can be introduced, and physical observables  are 
finite and independent of the regulator.  The 
Vacuum SFT (\ref{vsft}, \ref{QQ})
appears to capture the correct physics, at least at the classical level.
Taking a matter/ghost factorized ansatz
\be
| \Phi_g \rangle  \otimes  | \Phi_m \rangle \, ,
\ee
and assuming that the ghost part is universal for all D-branes solutions, 
the equations of motion
reduce to following equations for the matter
part,
\be \label{proj}
| \Phi_m \ra * | \Phi_m \ra = | \Phi_m \ra \,.
\ee
A solution  $| \Phi_m \ra$ can be regarded
as a {\it projector} acting in ``half-string space''.
 Recall that the $*$-product looks formally like a matrix multiplication (\ref{functional}):
  the matrices are the string fields, whose  ``indices'' run over the half-string curves.
These  projector equations have been exactly solved
by many different techniques (see (Rastelli, 2004) for a review).
In particular there is a general
BCFT construction that shows that one can obtain solutions corresponding to {\it any} D-brane
configuration,  including  {\it multiple} branes  -- the rank of the projector 
 is the number of branes. A rank one projector corresponds
 to an open string functional which is left/right {\it split}, $\Phi[X(\sigma)] = F_L(X_L) F_R(X_R)$.
 There is also clear analogy between these solutions
 and the soliton solutions  of non--commutative field theory.
 The analogy can be made  sharper
 using a formalism that re-writes
 the open string $*$-product as the tensor product of infinitely
 many Moyal products. (See (Bars, 2002) and references therein).

 It is unclear whether or not multiple brane solutions (should)  exist
 in the original OSFT -- they are yet to be found in level truncation.
  Understanding this and other issues, like
 the precise role of closed strings in the quantum theory (section 2.5)
 seems to require a precise
characterization of  the allowed space  of open string
functionals. In principle,  the path-integral
over such functionals 
would  define  the theory
at the full non-perturbative 
level.  This remains a challenge for the future.

\bigskip

{\it Note added in proof:}

\smallskip

Very recently, M. Schnabl, building on
previous work on star algebra projectors
and related surface states (Rastelli (2004) and references
therein) was able to find the exact solution
for the universal tachyon condensate in OSFT.
This breakthrough is likely to lead to rapid new developments in SFT.

\bigskip

{\bf  Acknowledgments}

It is a pleasure to thank Barton Zwiebach for  critical reading of the manuscript.

\bigskip

\bigskip

\bigskip

{\bf Further reading}

\bigskip

\bigskip

%\bibitem{Bars:2002yj}
 I.~Bars (2002), 
  ``MSFT: Moyal star formulation of string field theory,''
  arXiv:hep-th/0211238.
  %%CITATION = HEP-TH 0211238;%%

%\cite{Berkovits:2001nr}
%\bibitem{Berkovits:2001nr}
  N.~Berkovits (2001),
  ``Review of open superstring field theory,''
  arXiv:hep-th/0105230.
  %%CITATION = HEP-TH 0105230;%%

%\cite{Berkovits:2004xh}
%\bibitem{Berkovits:2004xh}
  N.~Berkovits, Y.~Okawa and B.~Zwiebach (2004),
  ``WZW-like action for heterotic string field theory,''
  JHEP {\bf 0411}, 038 
  [arXiv:hep-th/0409018].
  %%CITATION = HEP-TH 0409018;%%

%\cite{Erler:2004hv}
%\bibitem{Erler:2004hv}
  T.~G.~Erler and D.~J.~Gross (2004),
  ``Locality, causality, and an initial value formulation for open string field
  theory,''
  arXiv:hep-th/0406199.
  %%CITATION = HEP-TH 0406199;%%

S. B. Giddings, E. J. Martinec, E. Witten (1986) ``Modular invariance
in string field theory,''  Phys.\ Lett.\ B {\bf 176}, 362.
  %%CITATION = PHLTA,B176,362;%%

S. B. Giddings (1988), ``Conformal techniques in string field theory,''
 Phys.\ Rept.\  {\bf 170}, 167. 
  %%CITATION = PRPLC,170,167;%%

%\cite{Ohmori:2001am}
%\bibitem{Ohmori:2001am}
  K.~Ohmori (2001),
  ``A review on tachyon condensation in open string field theories,''
  arXiv:hep-th/0102085.
  %%CITATION = HEP-TH 0102085;%%
%\cite{Rastelli:2004cp}
%\bibitem{Rastelli:2004cp}

  %\cite{Okawa:2002pd}
%\bibitem{Okawa:2002pd}
  Y.~Okawa (2002),
  ``Open string states and D-brane tension from vacuum string field theory,''
  JHEP {\bf 0207}, 003 (2002)
  [arXiv:hep-th/0204012].
  %%CITATION = HEP-TH 0204012;%%

L.~Rastelli (2004)
  ``Open string fields and D-branes,''
  Fortsch.\ Phys.\  {\bf 52}, 302 (2004).
  %%CITATION = FPYKA,52,302;%%

%\cite{Rastelli:2001uv}
%\bibitem{Rastelli:2001uv}
  L.~Rastelli, A.~Sen and B.~Zwiebach (2001),
  ``Vacuum string field theory,''
  arXiv:hep-th/0106010.
  %%CITATION = HEP-TH 0106010;%%

M. Schnabl  (2005),   ``Analytic solution for tachyon condensation in open string field theory,''
  arXiv:hep-th/0511286.
   %%CITATION = HEP-TH 0511286;%%
  
%\cite{Sen:2004nf}
%\bibitem{Sen:2004nf}
  A.~Sen (2004)
  ``Tachyon dynamics in open string theory,''
  arXiv:hep-th/0410103.
  %%CITATION = HEP-TH 0410103;%%

%\cite{Siegel:1988yz}
%\bibitem{Siegel:1988yz}
  W.~Siegel (1988),
  ``Introduction To String Field Theory,''
  arXiv:hep-th/0107094.
  %%CITATION = HEP-TH 0107094;%%

%10)something on BSFT? Shatashvili's Mumbai talk?
%\cite{Shatashvili:2001ux}
%\bibitem{Shatashvili:2001ux}
  S.~L.~Shatashvili (2001),
  ``On field theory of open strings, tachyon condensation and closed
  strings,''
  arXiv:hep-th/0105076.
  %%CITATION = HEP-TH 0105076;%%

%\cite{Taylor:2003gn}
%\bibitem{Taylor:2003gn}
  W.~Taylor and B.~Zwiebach (2003),
  ``D-branes, tachyons, and string field theory,''
  arXiv:hep-th/0311017.
  %%CITATION = HEP-TH 0311017;%%

%\cite{Thorn:1988hm}
%\bibitem{Thorn:1988hm}
  C.~B.~Thorn (1989),
  ``String Field Theory,''
  Phys.\ Rept.\  {\bf 175}, 1 (1989).
  %%CITATION = PRPLC,175,1;%%

%\cite{Witten:1985cc}
%\bibitem{Witten:1985cc}
  E.~Witten (1986),
  ``Noncommutative Geometry And String Field Theory,''
  Nucl.\ Phys.\ B {\bf 268}, 253.
  %%CITATION = NUPHA,B268,253;%%

%\cite{Zwiebach:1992ie}
%\bibitem{Zwiebach:1992ie}
  B.~Zwiebach (1993),
  ``Closed string field theory: Quantum action and the B-V master equation,''
  Nucl.\ Phys.\ B {\bf 390}, 33 
  [arXiv:hep-th/9206084].
  %%CITATION = HEP-TH 9206084;%%

%\cite{Zwiebach:1997fe}
%\bibitem{Zwiebach:1997fe}
  B.~Zwiebach (1998),
  ``Oriented open-closed string theory revisited,''
  Annals Phys.\  {\bf 267}, 193 
  [arXiv:hep-th/9705241].
  %%CITATION = HEP-TH 9705241;%%

\end{document}